\documentclass[11pt,english]{article}
\usepackage{dolomieu,epsfig}
\def\Journal#1#2#3#4{{#1} {\bf #2}, #3 (#4)}

\def\be{\begin{equation}}
\def\ee{\end{equation}}
\def\bea{\begin{eqnarray}}
\def\eea{\end{eqnarray}}

\usepackage{babel} 
\begin{document}
\vspace*{4cm}

\title{DIRECT SEARCH FOR WIMP DARK MATTER}

\author{ J. GASCON }

\address{Institut de Physique Nucl\'eaire de Lyon, 4 rue Enrico Fermi\\
69622 Villeurbanne Cedex, France}

\maketitle\abstracts{
We will review the experimental aspects of the direct search
for WIMP dark matter.
In this search, one looks in a terrestrial target for nuclear recoils
produced by the impacts with WIMPs from the galactic halo.
After describing the different ingredients involved
in the calculation of rates in a given detector, we
will present the different search strategies and review
the currently running experiements and the prospects
of future experiments.
}


\section{Introduction}

One of the most exciting possibility opened by the recent cosmological
observations is that our Galaxy could be immersed in a halo of heavy
Dark Matter particles of a fundamentally new type. In the most likely
scenario these particles would be the WIMPs (acronym for Weakly Interacting
Massive Particles). The discussion on how the WIMP has come to be
one of the most actively sought hypothesis concerning Dark Matter
has been discussed in other lectures in this school. A detailed discussion
can be found in Ref.~\cite{jungman}. Here, we will review the experimental
aspects of direct WIMP search. In this search, one looks, in a terrestrial
target, for nuclear recoils produced by the impacts with WIMPs from
the galactic halo. After describing the different ingredients involved
in the calculation of rates in a given detector, we will present the
different search strategies and review the currently running experiments
and the prospects of future experiments.

As detailed discussions on this subject can be found in the literature,
this short text is only meant as an introduction to the subject. For
a more complete review on WIMP dark matter, we refer the reader to
Ref.~\cite{jungman}. A comprehensive description of the method to
interpret experimental WIMP search results can be found in Ref.~\cite{lewin}.

\section{Principles of Direct Detection}

\subsection{WIMP density}

Cosmological measurements, such as those of WMAP\cite{wmap}, provide
very strong incentives to look for dark matter. At these scales, the
density of dark matter is of the order of 1 GeV/c$^{2}$/m$^{3}$,
the equivalent in mass of one proton per cubic meter. In these measurements,
dark matter signals its presence via gravitational effects on ordinary
matter in the early universe. To clearly identify the nature of dark
matter, it is crucial to be able to observe non-gravitational interactions
with ordinary matter, or at least put an upper limit on the strength
of these interactions. Only then would we able to test one of the
most attractive scenario, namely that dark matter is made of Weakly
Interacting Massive Particles, where {}``weakly'' is meant in the
context of the nuclear weak force. Two types of search are proposed:
indirect or direct. Indirect searches look in cosmic rays for products
of annihilation of WIMP pairs. This is discussed in more details in
other lectures. Direct searches look for a nuclear recoil produced
by a collision with a WIMP from the halo of our galaxy. 

The rate of such collisions depend linearly on the local WIMP density
$\rho_{WIMP}$. A common estimate~\cite{lewin} for this quantity
is 0.3 GeV/c$^{2}$/cm$^{3}$. It should be noted that this local
density should not be confused with the cosmological density of dark
matter $\Omega_{DM}$, which is also approximately 0.3 when expressed
as a fraction of critical density of the universe. However the respective
units differ by more than a factor of $10^{6}$. One could have hoped
that the improved accuracy of the recent measurements of $\Omega_{DM}$
would reduce the uncertainties in WIMP searches. This is not exactly
the case, as $\rho_{WIMP}$ is not derived from $\Omega_{DM}$ but
from measurements of stars and gas in our galaxy, with cross-checks
based on the rotation curves of other galaxies. However the cosmological
measurements remain a strong incentive to look for local dark matter,
as it would be surprising if this main component of matter in the
universe remains absent of our neighborhood!

\subsection{WIMP velocity\label{sub:WIMP-velocity}}

\begin{figure}
\begin{center}\includegraphics[%
  scale=0.4]{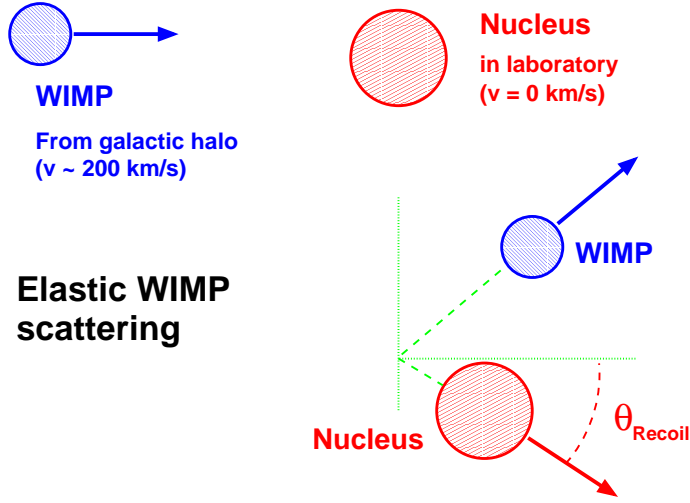}\end{center}

\caption{\label{cap:Kinematics}Kinematics of a WIMP-nucleus collision.}
\end{figure}

Assuming a local WIMP density of 0.3 GeV/cm$^{2}$, another ingredient
is needed to estimate the WIMP flux in this room: their velocity distribution
$f(v)$. The order of magnitude of their average velocity is fixed
by the assumption that the halo WIMPs are gravitationnally bound to
the galaxy (and its halo), leading to velocities of the order of stellar
velocities in our galaxy, approximately 200 km/s. We will come back
later on a more accurate description. For now, we can make the following
order-of-magnitude calculation. For WIMPs with masses of approximately
100 GeV/c$^{2}$(the mass of a A=100 nucleus, we will see later the
motivation for this example), the local density is 3000 WIMP per cubic
meter, and a flux of $6\times10^{4}$ WIMPs is traversing each cm$^{2}$
of our body every second. Another important aspect is that the average
kinetic energy of these WIMPs is 20 keV. This energy is much larger
than the $\sim$eV scale binding energy of nuclei in a solid. In direct
searches, the collisions are detected by the measurement of the energy
of the recoiling nucleus, as its kinetic energy is dissipated in the
detector medium. The simple kinematics of the collision is shown in
Fig.\ref{cap:Kinematics}. In such a collision, the energy of the
nuclear recoil $E_{recoil}$ is given by:

\begin{equation}
E_{recoil}=E_{WIMP}\frac{{4M_{nucleus}M_{WIMP}}}{(M_{nucleus}+M_{WIMP})^{2}}\cos^{2}\theta_{recoil}\label{eq:erecoil}\end{equation}
where $E_{WIMP}$is the initial kinetic energy of the WIMP, $M_{WIMP}$
is its mass, $M_{nucleus}$ is the mass of the recoiling nucleus and
$\theta_{recoil}$ is the angle of the nuclear recoil relative to
the initial WIMP direction. From this equation, we can derive that
the maximal recoil energy is obtained when $M_{WIMP}=M_{nucleus}$.
This search is thus more efficient for a WIMP with a mass comparable
to nuclear masses.

As the resulting recoil energy depends on the kinetic energy distribution
of the WIMPs in our halo, a more precise estimate of this velocity
distribution is needed. Unfortunately, it is extremely difficult to
calculate this distribution, even if we arbitrarily fix the total
mass of the halo and neglect interactions with ordinary matter and
non-gravitational interactions. The reason is that the gravitational
force has an infinite range and the number of WIMPs to be included
in the calculation is extremely large. The simplest case (see e.g.
Ref.~\cite{lewin}) is to assume that the halo is spherical and that
the WIMPs trapped in the galactic field have attained thermal equilibrium,
with a Maxwellian velocity distribution:\begin{equation}
\frac{dP(v)}{v^{2}dv}=\frac{1}{(\pi v_{0}^{2})^{3/2}}\exp({-\frac{v^{2}}{v_{0}^{2}}})\label{eq:max}\end{equation}
where $v_{0}\sim$220 km/s ($v_{rms}=\sqrt{{\frac{{3}}{2}}}v_{0}=$270
km/s). To be consistent, this distribution is generally truncated
at the velocity at which a WIMP could escape the galaxy ($v_{escape}\sim$650
km/s). This model has many known shortcomings. It predicts a steep
increase in density at the core of the galaxy that is not supported
by observations. Many-body calculations also tend to produce non-uniform
spatial density distributions (with so-called {}``clumps'' of dark
matter), with strong non-uniformity effects in phase space ({}``caustics'').
In addition, the halo may not adopt a spherical shape (for example,
it could be triaxial), and tidal flow from neighboring galaxies may
play an important role. In view of these difficulties and in the absence
of a consensus, the isothermal Maxwellian distribution is generally
adopted for the analysis of dark matter searches. This is acceptable,
as most searches are only sensitive to the average WIMP kinetic energy.
This may evolve once a signal with significant statistics is observed.

\begin{figure}
\begin{center}\includegraphics[%
  scale=0.8]{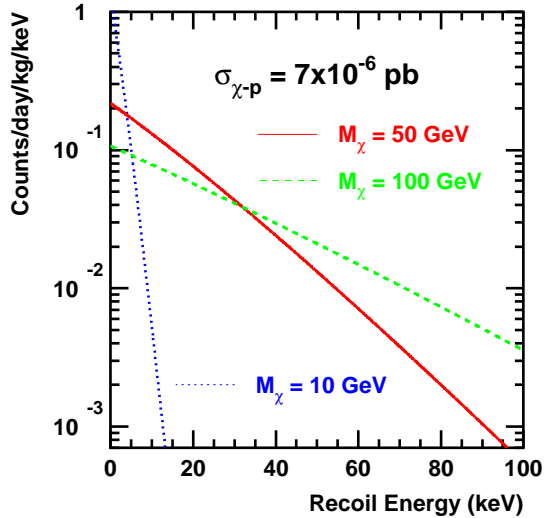}\end{center}

\caption{\label{cap:Erecoil-vs-M}Recoil energy distributions for three different
WIMP masses in germanium.}
\end{figure}

With the simple Maxwellian distribution (eq.\ref{eq:max}), the relation
between $E_{recoil}$ and $E_{WIMP}$ (eq.\ref{eq:erecoil}) and the
assumption of an isotropic $\theta_{recoil}$ distribution in the
center-of-mass system of the collision, it is easy to derive the shape
of the recoil energy distribution:\[
\frac{{dN}}{dE_{recoil}}\propto\exp(-E_{recoil}/<E_{recoil}>)\]
where\begin{equation}
<E_{recoil}>=\frac{{\mu^{2}v_{WIMP}^{2}}}{M_{recoil}}\label{eq:avgErec}\end{equation}
where $v_{WIMP}^{2}$ is the average square velocity of the WIMP and
$\mu$ is the reduced mass of the WIMP-nucleon system. Taking into
account the finite escape velocity and the effects discussed in the
following will not alter significantly this overall shape. Example
distributions for different WIMP masses are shown for a germanium
target in fig.\ref{cap:Erecoil-vs-M}%

\begin{figure}
\begin{center}\includegraphics[%
  scale=0.6]{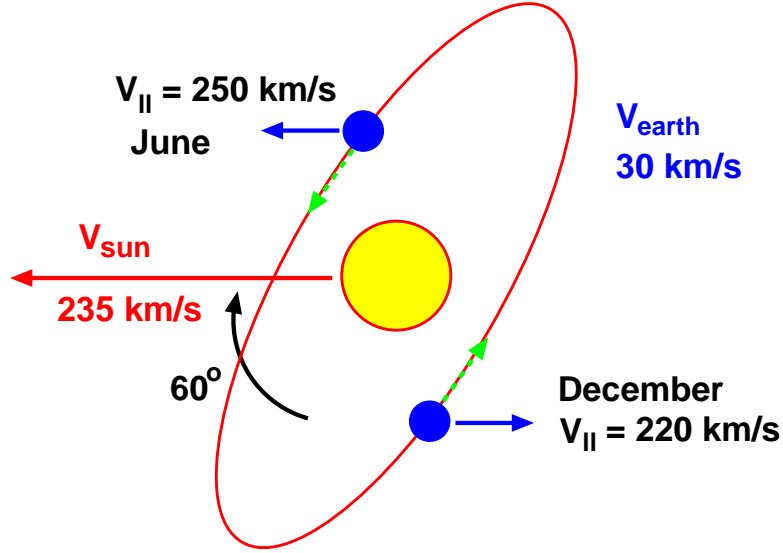}\end{center}

\caption{\label{cap:Earth-and-sun}Earth and sun tangential velocity relative
to the center of the Galaxy}
\end{figure}

A feature of the velocity distribution that is known with precision
and can be easily implemented is the fact that the nuclear targets
are not at rest relative to the galaxy, but are following the earth
and the sun motions (fig.\ref{cap:Earth-and-sun}).%
 The tangential velocity of the sun around the galactic center (in
the direction of Sagittarius) is 235 km/s. The net effect is to boost
the average kinetic energy of the WIMP flux on earth. The earth velocity
is an order of magnitude smaller (30 km/s) and can generally be neglected,
except for an interesting modulation effect in the flux. As the earth
orbits the sun with a 60$^{o}$angle relative to the galactic plane,
a 30$\times\cos60^{o}$=15 km/s velocity component is alternatively
added and subtracted to the sun's velocity relative to the WIMP flux.
This may result in a $\pm$7\% annual modulation of the collision
rate that can provide an interesting experimental signature of the
astrophysical nature of the detected signal, although recent calculations
have shown that the actual size of the effect may depend a lot on
the details of the halo models (see e.g. refs.\cite{copi,green}).

To really make an efficient use of the modulation effects, a more
powerful approach would be to detect for each recoil not only its
kinetic energy $E_{recoil}$ but also its direction $\theta_{recoil}$.
The apparent direction of the WIMP flux should be correlated with
that of the motion of the sun around the galaxy, and have, in addition
to the annual modulation, a diurnal modulation as the laboratory follows
the earth rotation on its axis. However it is difficult to measure$\theta_{recoil}$,
as the typical recoil range of a recoil is of the order of 20 nm in
a crystal (for a 20 keV Ge recoil in Ge, for example) and 30 $\mu$m
in a gas (for a 20 keV Kr recoil in Kr). Detectors aiming at measuring
the spatial extension of such short tracks are still at an early R\&D
stage (see e.g. DRIFT\cite{drift}). So far, the most sensitive experiments
measure only $E_{recoil}$ and are blind to the recoil direction.

\subsection{Recoil Energy Measurement}

Now comes the question of the method to detect the recoil energy.
A nucleus with 20 keV kinetic energy will dissipate this energy in
a crystal via three main processes: ionization, scintillation and
phonons. The ionization corresponds to the electrons stripped by the
initial nucleus and the following cascade. In certain material, this
electronic activity will emit scintillation light. The movement of
the incident nucleus in the lattice will also induce vibrational phonons.
In a closed system, all ionization and scintillation energy will convert
into phonons that will eventually thermalize and produce an elevation
of temperature of the crystal.

In nuclear physics, the most common technique to detect radiation
in the 20 keV range is the use of scintillating crystals (such as
NaI or BGO) and solid state semiconductor ionization detectors (Ge
or Si). However, these detectors are usually optimized for gamma-ray
radiation and not nuclear recoils. When a keV to MeV range photon
enters an detector, it converts most of its energy to an electron.
The range of this electron is of the order of the $\mu$m , much greater
than the nm range of nuclear recoils, and as a consequence, this electron
will produce more ionization than a nuclear recoil of equal energy,
as the latter will lose a substantial part of its energy directly
into phonons associated with atom vibrations as the nucleus is stopped
in the lattice. Scintillation yields of electron and nuclear recoils
will be similarly affected. A quantitative measure of this effect
is the so-called \emph{quenching factor} (symbol: \emph{Q}). Its definition
is the following. First, the ionization, scintillation or heat response
of a detector to gamma-ray of known energy is measured, yielding a
calibration of the signal in keV-equivalent-electron (keV$_{ee}$).
Then, the response of the detector to a nuclear recoil of known energy
$E_{recoil}$ is measured. As the ionization or scintillation yield
will be lower than in the electron recoil case, the measured signal
in keV$_{ee}$ will be $E_{ee}=QE_{recoil}$, where $Q$ is a fraction.
With this definition, $Q$ is the relative signal yield for nuclear
and electron recoils. For a recent review of quenching measurements,
see Ref.\cite{sicane} and references therein. For the heat measurement
in perfectly isolated detectors, $Q$ should be unity, if one waits
long enough for the complete thermalization of all energy. In ionization
detectors such as Ge or Si, $Q\sim0.3$, with a moderate energy dependence.
For scintillation, there is a wider range of values: in NaI, the quenching
for Na and I recoils are $\sim0.25$ and$\sim0.09$, respectively.
Another noteworthy case is scintillation in Xe ($Q\sim0.2$). The
quenching factor must be kept in mind when comparing different detector
results, as the measured energies are often quoted in keV$_{ee}$
instead of true recoil energy. More interestingly, as will be shown
later, this effect can be put to contribution as a mean to discriminate
nuclear recoils from the usually large background of electronic recoils.

\subsection{Order-of-magnitudes of scattering cross-sections}

In order to estimate the rate of collisions between WIMP and nucleons,
one needs to define which elementary force mediates these encounters.
Gravitational interactions between a single WIMP and a single nucleus
are negligible. Electromagnetic interactions are excluded, since it
would mean that WIMP could emit or absorb light. However, it has been
observed that the behavior of the dark matter particles throughout
the Big Bang up to now could be explained by simply assuming that
they only participate to weak interactions. If this is the case, this
leads to an estimate of the probability of a collision with a nucleus.

In particle and nuclear physics, the probability of an interaction
is usually expressed as deriving from a \emph{cross}-\emph{section},
$\sigma$, with units of surface. If $dN/dt$ is the number of WIMP-nucleus
interactions per unit time, $\phi$ is the WIMP flux and the number
of target nuclei per volume, we have\[
\frac{{dN}}{dt}=\phi\sigma_{A}N_{target}\]
where $\sigma_{A}$ is the cross-section for a WIMP-nucleus collision.
A typical cross-section for a collision on a $A\sim100$ nucleus involving
the nuclear force only is of the order of the size of this nucleus:
$10^{-24}$cm$^{2}$ = 1 barn (symbol: b). Collisions involving electromagnetic
interactions will have a much larger cross-section, due to the long-range
nature of the force. If the nuclear Weak force is involved, the cross-section
is at most 1 picobarn (1 pb = 10$^{-12}$b). Typical weak cross-sections
on single nucleon (a proton or a neutron) are even lower than this
($\sigma_{n}\sim10^{-7}$pb). With such cross-sections, the interaction
rate with the WIMP flux can be expected to be at most one collision
per kilogram of matter per day, possibly as low as one per year and
per ton of detector.

\subsection{Scaling from a nucleon to a nucleus}

The reason why the weak cross-section on a $A=100$ nucleus is not
simply 100 times that on a single nucleon is that the wavelet associated
to the momentum transfer corresponding to a A=100 nucleus with 20
keV kinetic energy is approximately 3 fm, about the size of the entire
nucleus. In this case, one must evaluate whether the interaction goes
through a spin-dependent or scalar (spin independent) process. In
the first case, only the unpaired nucleon will contribute significantly
to the interaction, as the spins of the $A$ nucleon in a nucleus
are systematically anti-aligned. In the second case, all nucleon contributions
add coherently: the total amplitude scales as $A$ and the total scattering
probability as $A^{2}$. Another mass-dependence hidden in the scaling
from $\sigma_{n}$ to $\sigma_{A}$ is that interaction probability
depends on the density of states in the final state, which in this
case\cite{Perkins}, implies that $\sigma_{A}/\sigma_{n}=\mu_{A}^{2}/\mu_{n}^{2}$,
where $\mu_{A}$ ($\mu_{n}$) is the invariant mass of the WIMP-nucleus
(WIMP-nucleon) system. In summary, the $A-$dependence of WIMP-nucleus
cross-section is:

\[
\sigma_{A}=\frac{{\mu_{A}^{2}}}{\mu_{n}^{2}}\,\sigma_{n}\, A^{2}\mbox{{\,(spin independent)}}\]

\[
\sigma_{A}=\frac{{\mu_{A}^{2}}}{\mu_{n}^{2}}\,\sigma_{n}C\, J(J+1)\mbox{{\,(spin dependent)}}\]

where $C$ is a factor that depends on the details of the structure
of the nucleus\cite{lewin}. It cannot be expressed in a simple form,
but is generally less than unity.

As $\mu_{A}^{2}/\mu_{n}^{2}\sim A^{2}$ , the interaction rate per
kilogram of target mass is proportional to $A^{3}$in the case of
spin-independent interactions and only to $A$ in the case of spin-dependent
interactions. Direct searches try to benefit from this scaling by
using targets with as large $A$ as possible. In any model where some
part of the interaction involves spin-independent interactions, this
term dominates the cross-section.

\subsection{Nuclear form factors}

It was stated that the advantageous $A^{2}$ scaling of the spin-independent
cross-section arises from the fact that the wavelength associated
with the momentum transfer is comparable to the size of the nucleus.
To be more precise, full coherence is only achieved when the associated
wavelength is much larger than the nucleus size. In the present case,
one has to take into account interference effects that can be calculated
rather precisely using the known form factors. The nuclear structure
models behind these calculations and the size of the effect are discussed
in Ref.\cite{lewin}; here it suffices to say that the net effect
in most commonly used target material is to reduce the interaction
rate by a factor of 2 to 4, which damps the increase due to the $A^{2}$dependence
when $A\sim100$.

\subsection{Supersymmetric models predictions}

With precise prescriptions on the choice of $\rho_{WIMP}$, $f(v)$,
the $\sigma_{A}/\sigma_{n}$scaling and the nuclear form factor, the
only two missing ingredients for predicting the WIMP rate in a given
detector are $M_{W}$ and $\sigma_{n}$. For these, one needs a model
with specific predictions on the nature of the WIMP. 

\begin{figure}
\begin{center}\includegraphics[%
  scale=0.6]{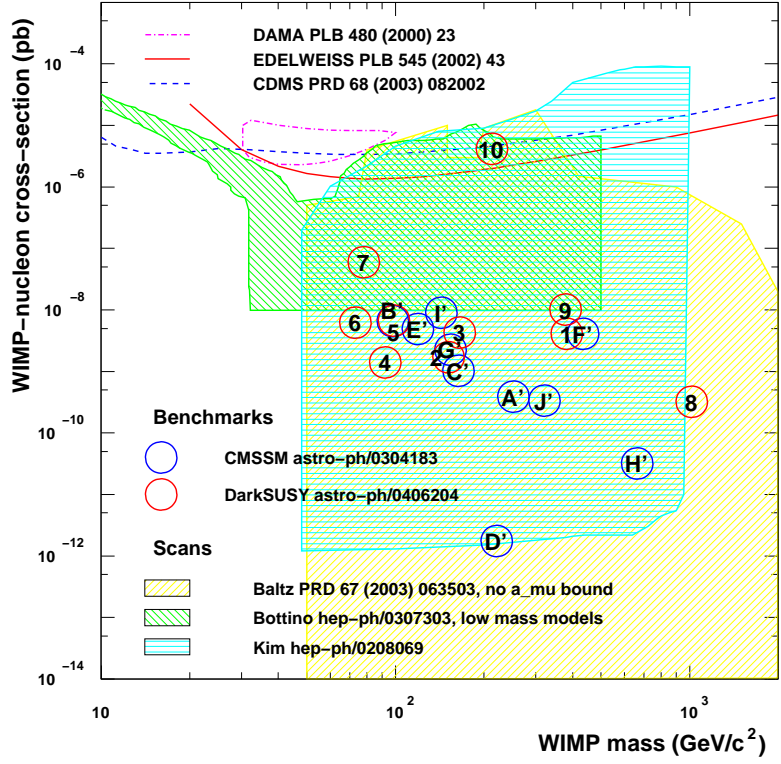}\end{center}

\caption{\label{cap:susypred} Predicted ranges of $M_{W}$ and $\sigma_{n}$
in different supersymmetric models. Published sensitivities of some
experiments are shown as a comparison. See text for references.}
\end{figure}

More precise predictions on the interaction rates can be obtained
within the framework of Supersymmetry (SUSY). In fact, this theory
actually predicts very naturally that there exists a heavy neutral
particle (called the neutralino, with symbol $\chi^{0}$) with weak
interaction only, that was created copiously at the Big Bang. And
this, quite remarkably, even though the theory was not motivated as
to fix the dark matter problem, but to solve basic problems in the
quantum description of the behavior of elementary particles. The great
advantage of this model is that its predictions are somewhat constrained
by the results of searches for supersymmetric particles and for deviations
from the Standard Model of interactions. For example\cite{jungman},
the mass of the neutralino cannot be much greater than 1 TeV/c$^{2}$.
It cannot be less than 50 GeV/c$^{2}$, except in exceptional versions
of the model that are specially tuned for this purpose\cite{bottino}.
In addition, models with purely spin-dependent interactions are essentially
ruled out, and in this framework, large-$A$ detectors are clearly
favored. Fig.\ref{cap:susypred} shows the range of $M_{W}$ and $\sigma_{n}$
allowed by different versions of SUSY models. In addition to these
so-called {}``scans'' of different versions\cite{bottino,baltz,kim}
of the Minimum Supersymmetric Model (MSSM), other calculations\cite{scan}
offer {}``bench mark models'' that explore interesting cases. Typical
values of $\sigma_{n}$ are in a $10^{-11}$ to $10^{-7}$ pb. This
is a few order of magnitudes below the sensitivity achieved by current
detectors\cite{dama,edw2002,cdms1}, also shown in Fig.\ref{cap:susypred}.

\section{Search Strategies}

\subsection{Search sensitivities}

\begin{figure}
\begin{center}\includegraphics[%
  scale=0.6]{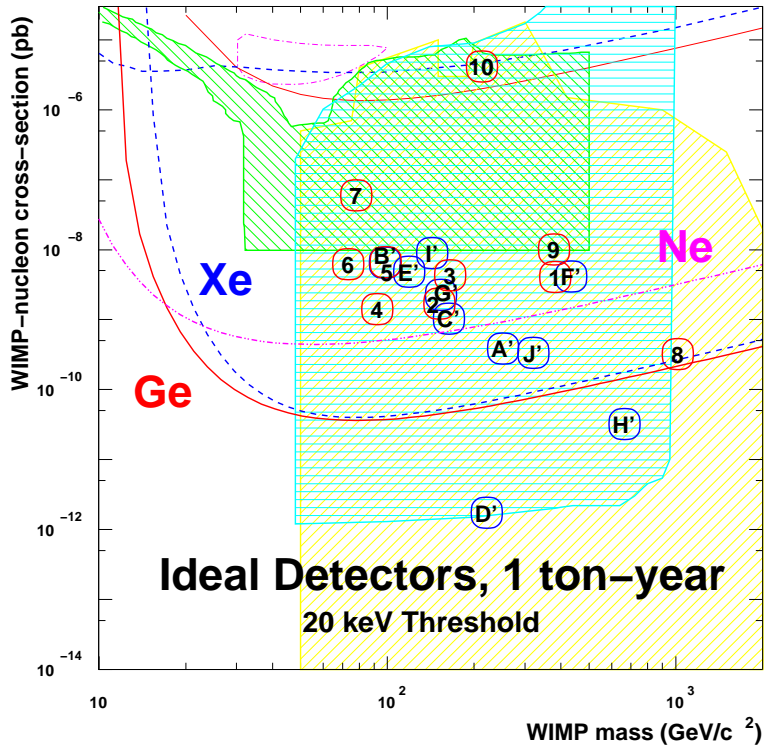}\end{center}

\caption{\label{cap:oneton}Sensitivity of a one-ton size detector running
over a full year with perfect background rejection. The supersymmetric
predictions are those of Fig.\ref{cap:susypred}.}
\end{figure}

However, as it can be seen in Fig.\ref{cap:oneton}, a sensitivity
of the order of $10^{-10}$pb is within the reach of a one-ton size
detector running over a full year with a perfect rejection of background,
and with a recoil energy threshold of 20 keV. Detectors with large-$A$
targets are favored over light-$A$ target such as Ne ($A=20$). As
described in the previous section, increasing $A$ from $^{73}$Ge
to $^{131}$Xe does not improves the sensitivity because of the evolution
of the nuclear form factor and also because of the $1/M_{recoil}$
dependence of the average recoil energy (Eq.\ref{eq:avgErec}). As
expected from kinematics, the experiments are the most sensitive for
$M_{W}\sim M_{nucleus}$. At lower masses, the presence of a fixed
recoil energy threshold further deteriorates the performance of the
search.

For a $M_{W}=100$ GeV/c$^{2}$, a cross-section of $10^{-6}$ pb
correspond to approximately 0.1 collisions per kg per day (kg$\cdot$d);
cross-sections of $10^{-8}$ and $10^{-10}$ pb corresponds respectively
to 0.5 events per kg per year (kg$\cdot$y), and 5 per ton and per
year (t$\cdot$y). 

The sensitivities of these ideal experiments are calculated the following
way. If no events are observed in the ton of target over the year,
an upper limit with a 90\% confidence level can be ascribed to $\sigma_{n}$.
Lets call this limit $\sigma_{n}^{90\%}$. This 90\% confidence level
(C.L.) means that in the case that $\sigma_{n}=\sigma_{n}^{90\%}$,
the probability of having a one ton-year experiment with zero observed
events is 10\%. In the 90\% other cases, the number of observed events
would be 1, 2 or more. The probability distribution of this background-less
experiment is given by the Poisson distribution\cite{statistics}:

\[
P(n,\mu)=\frac{{\mu^{n}}}{n!}\exp(-\mu)\]

where $P(n,\mu)$ is the probability of observing $n$ events in a
random process where the average number of observed events is $\mu$.
$P(0,\mu)=0.9$ correspond to $\mu=-\ln(1-0.9)=2.30$, so that $\sigma_{n}^{90\%}$corresponds
to the cross-section that would yield on average 2.30 events in the
detector. The lines drawn on Fig.\ref{cap:oneton} correspond to $\sigma_{n}^{90\%}(M_{W})$.

In the presence of a background, the sensitivity deteriorates significantly.
In the limit that the number of observed background counts $n_{bkg}$
is large, the sensitivity as measured by $\sigma_{n}^{90\%}$ becomes
equal to the cross-section yielding on average $n_{bkg}$ events in
the detector. As the time increases, so does $n_{bkg}$ and an increased
exposure does not yield a better sensitivity. Even if the background
can be evaluated precisely by an independent measurement and subtracted
from the number of observed events $n$, the statistical Poisson fluctuations
on $n$ remains. In this case the sensitivity $\sigma_{n}^{90\%}$
corresponds to the cross-section predicting an average number of observed
signal events of $1.28\sqrt{{n_{bkg}}}$, if $n_{bkg}<\sim20$. As
the exposure increases, the sensitivity grows very slowly with time
($\sqrt{{t}}$).

\subsection{Low backgrounds}

Extremely low background levels are thus essential for reaching sensitivities
covering the range of the MSSM predictions. As a comparison, the radioactivity
of a human body represents $10^{7}$ decays per kg$\cdot$d, with
most of them depositing more than 100 keV of energy. This is very
far from the 0.1 decays per kg$\cdot$d necessary to achieve a $10^{-6}$
pb sensitivity. The background from natural radioactivity has two
sources: external and internal radioactivity.

The shielding from external sources of radioactivity is achieved by
surrounding the detector with thick walls of absorbing material. A
high-Z material like lead is very effective for stopping MeV-energy
gamma-rays, while a few mm of low-Z material are sufficient for stopping
low-energy gamma-rays as well as beta and alpha radiations. Beyond
a thickness of 15 to 25 cm of lead, one is generally limited by the
internal radioactivity of lead itself. Fast neutrons are not by number
a large part of natural radioactivity, but they are of concern in
direct searches, as they produce nuclear recoils similar to those
produced in WIMP collisions. Fast neutron shields consist moderators
made of material with a high density of hydrogen, such as polyethylene
or water. 

\begin{figure}
\begin{center}\includegraphics[%
  scale=0.6]{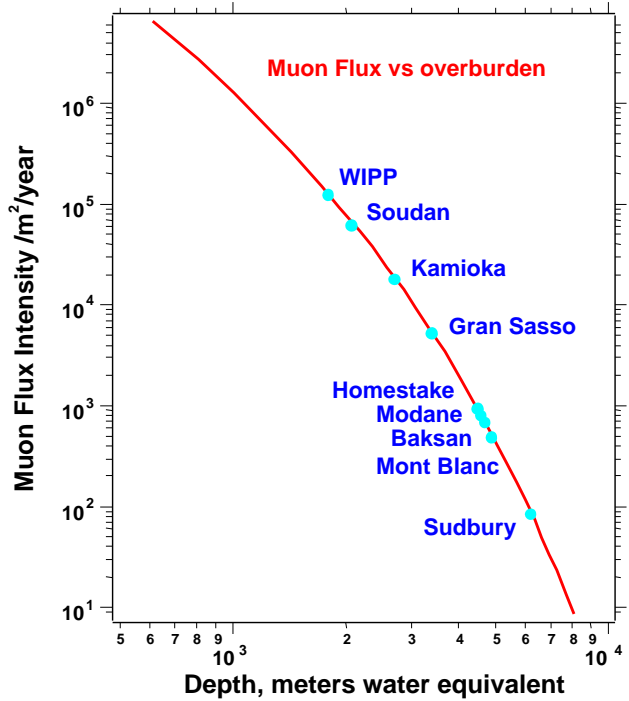}\end{center}

\caption{\label{cap:muon}Muon flux in muons per m$^{2}$per year and per
steradian in different underground laboratories.}
\end{figure}

Good internal radioactivity is achieved by using detectors made of
radiopure material. This limits the choice of detector technology.
In addition, it is necessary to place the detector in a deep-underground
site, where it is protected from the penetrating cosmic muon flux.
At ground level, this radiation (approximately $10^{3}$ muons per
cm$^{2}$ per day) induces nuclear transmutations to unstable isotopes
throughout the detector volume. In underground laboratories such as
Soudan Mine in Minnesota, the Gran Sasso laboratory in Italy or the
Laboratoire Souterrain de Modane in the Fr\'ejus Tunnel, this flux
is reduced by a factor varying from $10^{5}$to $10^{7}$(see Fig.\ref{cap:muon}).

\subsection{WIMP signatures}

To these passive shielding setup, one generally adds active background
rejection techniques, where an energy deposit due to a non-WIMP source
is identified by their different signatures. The two extreme cases
of background rejection are the \emph{event-by-event rejection}, where
each energy deposit in the detector is associated to an additional
signal that can be used to reject background events with a 100\% certainty,
and the \emph{statistical rejection}, where the additional information
can be used to ascertain which fraction of the total event sample
comes from a well-defined type of background, but cannot tell for
one individual event. In the second case, the precision on the WIMP
rate is limited by the statistical fluctuation on the total sample,
background included. The first case is ideal, as in practice there
is always a small probability that a background may fake the signature
of a WIMP. However, if this probability is small and the expected
number of fake WIMPs is less than one, the rejection can be considered
to be truly made event-by-event. Available WIMP signatures are:

\textbf{\emph{Nuclear recoils:}} WIMPs produce nuclear recoils, while
most radioactive backgrounds interact via the electromagnetic force
and produce electron recoils. The discrimination of nuclear and electron
recoils is generally based on the fact that the former have a larger
energy loss per unit length (dE/dx) and a smaller recoil range. This
also leads to the previously discussed quenching effects, as well
as difference in scintillation time constants in some crystals. 

\textbf{\emph{The shape of the recoil energy spectrum:}} The shape
of the $E_{recoil}$ spectrum for a WIMP with a given mass can be
calculated rather precisely. The observed energy spectrum must be
consistent with the expectation. Even if few events are expected,
the predicted spectrum shape is a useful tool to define the optimal
$E_{recoil}$ search range, which may vary as a function of $M_{W}$.
However, the overall shape is exponential, as is the case for many
background sources.

\textbf{\emph{Coherence:}} For spin-independent interaction, the scattering
cross-section should be proportional to $\mu^{2}A^{2}$. The observed
rates in different detectors should obey this law. For example, the
scattering cross-section of fast neutrons is approximately equal to
the geometrical cross-section of the target nucleus, and corresponds
to a $A^{2/3}$ dependence. 

\textbf{\emph{Multiple interactions:}} The mean free path of a WIMP
in matter is of the order of a light-year, so the probability of two
consecutive interactions in a single detector or two adjacent detectors
is completely negligible. In comparison the mean free path of a high
energy gamma-ray or a neutron is of the order of the cm and multiple
interactions are more common. An array of closely packed detectors
can efficiently identify these backgrounds.

\textbf{\emph{Uniform rate throughout the detector:}} The long mean
free path of WIMPs also means that their interactions must be spread
evenly throughout the detector volume. If the detector size is significantly
larger than the mean free path of high-energy photons or neutrons,
the interaction of the radiation originating from the surrounding
material and surface contaminant will occur mostly at the detector
surface. This leads to the incentive of building large position-sensitive
detectors. Other type of radiation have very short mean free path
($<$mm), such as low-energy photons, beta and alpha rays, and can
be rejected even if the position sensitivity is limited to the identification
of energy deposits located near the surface of the detector.

\textbf{\emph{Annual modulation:}} As discussed in section\ref{sub:WIMP-velocity},
the WIMP flux and its average kinetic energy modulates annually as
the earth alternatively adds and subtracts its velocity to the sun
movement relative to the galaxy. In the absence of annually-modulated
backgrounds, this behavior may be used as a WIMP signature, although
the size of the modulation is more dependent on the details of the
halo model than the year-averaged rate. Another drawback is that the
statistical uncertainty on the modulated signal is dominated by the
contribution of the large non-modulated component of the WIMP rate.
For example, even in the total absence of background, a 3$\sigma$
measurement (33\% relative error) of a $\pm$2.5\% modulation of the
rate requires a sample of at least 36000 WIMPs. Current background-free
experiments are excluding at 90\% C.L. rates corresponding to 0.2
events per kg$\cdot$d, with exposures of the order of 10 kg$\cdot$d.
Consequently, the observation of a 3$\sigma$ modulation effect would
require an exposure greater than half a ton-year!

\textbf{\emph{Directionality:}} see section \ref{sub:WIMP-velocity}.

In order to be convincing, an eventual WIMP signal should combine
more than one of these signatures. In addition, the results of direct
searches should be compatible with those of indirect searches, and
with the properties of the type of neutralino that would be eventually
discovered at the Large Hadron Collider.

\subsection{Current status of direct searches}

\begin{figure}
\begin{center}\includegraphics[%
  scale=0.7]{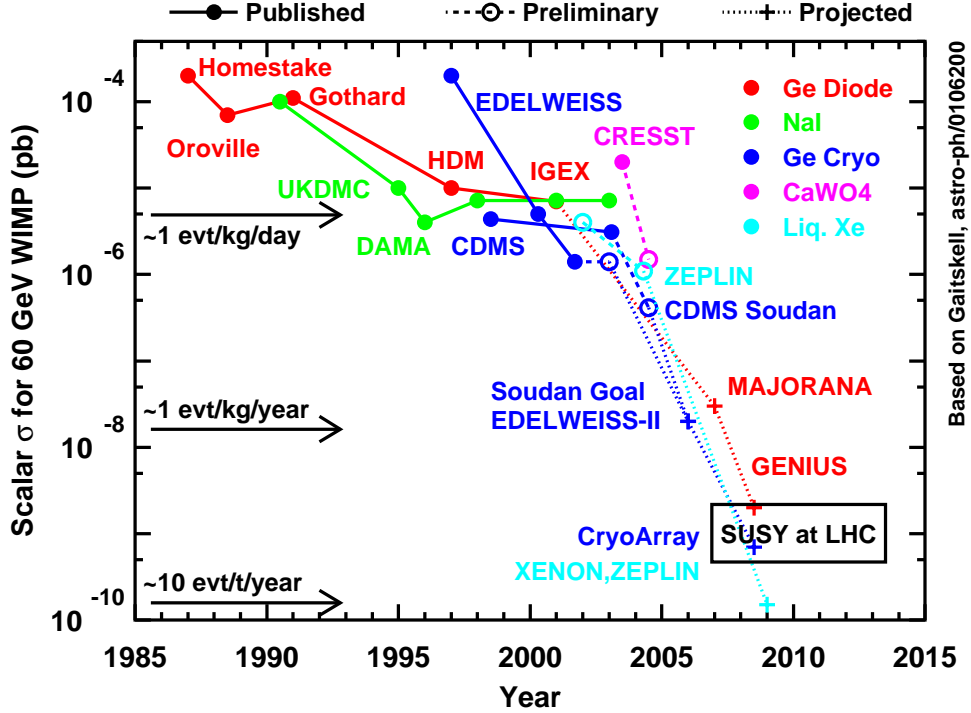}\end{center}

\caption{\label{cap:Status-and-evolution}Status and evolution of current
direct searches experiments. See text for explanation.}
\end{figure}

Fig.\ref{cap:Status-and-evolution} shows the evolution as a function
of time of the sensitivity for a 60 GeV/c$^{2}$ WIMP of different
direct searches. The different detector techniques will be described
in turn in the next section. Published results are shown as full symbols
and lines. Preliminary results are shown as open symbols and dashed
lines, while projections and estimates are represented with crosses
and dotted lines. At present, the most competitive direct searches
have reached sensitivities close to $10^{-6}$ pb. This starts to
explore the domain of optimistic Supersymmetric models. In the coming
years, different projects are planning to reach sensitivities approaching
$10^{-8}$ pb. This represents a factor 100 increase in performance.
This phase should help identify which technologies are suited for
the next ambitious goal, to achieve a further factor 100 in performance
and reach $10^{-10}$ pb-scale sensitivity with ton-scale detector
arrays. Only then will we be able to cover most of the predictions
of supersymmetry. 

\section{Review of Present Experiments}

\subsection{Ionization}

Already used in the first early WIMP searches, Ge ionization detectors
benefit from the high intrinsic purities achieved by the semiconductor
industries and from the developments in the context of the search
of the neutrino-less double-$\beta$ ($0\nu2\beta$) decay of $^{76}$Ge.
Despite the impressively low raw rates obtained by the Heidelberg-Moscow\cite{hdm}
and IGEX\cite{igex} experiments, the lack of event-by-event rejection
of electronic recoils means that the sensitivity of this technique,
now limited to approximately $\sim$1 event per kg$\cdot$d for WIMP
searches, can only be improved by further efforts on the radiopurity
of the detector environment and by the exploitation of the self-shielding
possibilities offered by large and compact arrays of detectors. With
a strong motivation in the context of $0\nu2\beta$ decay searches,
two ton-scale arrays are being developed: GENIUS\cite{genius}, with
naked Ge detectors immersed in liquid nitrogen, and MAJORANA\cite{majorana}.
This project is developping detectors with highly segmented electrodes,
with the intent of identifying multi-hit events by the study of the
shape of the pulses on the different segments. This technique is well
suited for the rejection of the multiple Compton scattering events
produced by the$\sim$2 MeV photons which constitute an important
source of background in the $0\nu2\beta$ decay searches. More studies
are needed to evaluate its suitability to the low-energy signals associated
with WIMP-induced nuclear recoils.

\subsection{Scintillating Crystals}

Scintillating crystals like sodium iodine (NaI) are a convenient solution
to accumulate large masses on detector material. It is however more
difficult to achieve radiopurity comparable to Ge. NaI-based searches,
such as DAMA\cite{dama}, ELEGANT\cite{elegant} or NAIAD\cite{naiad},
originally attempted to use pulse shape discrimination to statistically
identify a WIMP component in their observed rate. It was found that
the low number of detected scintillation photon per keV of incident
energy ({}``photo-electron per keV'', or p.e./keV$_{ee}$) restricts
the usefulness of this method at low energy. The technique is now
being investigated for CsI scintillator\cite{csi}, where the difference
in time constants between electron- and nuclear-recoil induced scintillation
is larger than in NaI.

The limitation of pulse shape analysis at low energy enticed the DAMA
collaboration to turn to a statistical discrimination based on annual
modulation\cite{dama}. With a data set of $10^{5}$ kg$\cdot$d recorded
with a 100 kg array of NaI over eight years, DAMA reports the observation
of a modulation originally interpreted as a WIMP with $M_{W}$ = 52
GeV/c$^{2}$ and $\sigma_{n}$= 7.2$\times$10$^{6}$pb. Such a WIMP
corresponds to a total rate of approximately 1 nuclear recoil per
kg$\cdot$d above a threshold of 2~keV$_{ee}$ corresponding to approximately
22 keV recoil energy. Reconciliating the reported modulation effect
with the published exclusion limits based on cryogenics Ge detectors\cite{edw2002,cdms1},
and with indirect search results requires very strong excursions from
the usual supersymmetric neutralino scenario\cite{copi,apan}. DAMA
is nevertheless planning an upgrade to a larger mass of detectors
(LIBRA).

\subsection{Noble Liquids}

Noble liquid scintillating detectors can provide large volume of highly
purified target material. Xenon\cite{zeplin,xenon,xmass} is particularly
interesting because of its high-$A$ value, but argon\cite{warp}
and neon\cite{clean} are also studied. Despite the development of
very efficient purification techniques developed for noble gases (originally,
to improve the stability of the scintillation properties), surface
radioactivity from the liquid container and the signal extraction
system is to be expected and some event-by-event background rejection
is required.

For xenon, different strategies are being investigated. The ZEPLIN
collaboration\cite{zeplin} has reported results from pulse shape
discrimination based on the different scintillating time constants
of nuclear and electron recoils in a 6 kg liquid Xe cell. However,
it is now moving to a two-phase (liquid+gas) detector, where the main
discrimination comes from the difference in ionization and scintillation
yields of electron and nuclear recoil events. A strong electric field
drifts the ionized electrons out of the liquid phase into the gas
phase, where they are detected via the secondary luminescence. The
collaboration XENON\cite{xenon} is preparing a 100~kg two-phase
project, with nuclear/electron recoil discrimination coming from the
ionization/scintillation yields, and will exploit the position-dependence
of the signals to define a fiducial volume away from surface contaminations.
Another Xe project is XMASS\cite{xmass}, where the emphasis is put
on position resolution in order to reject events due to surface contamination
and multiple scattering inside the 100-kg detector.

\subsection{Bolometers with Discrimination}

With this type of detector, the emphasis is on event-by-event rejection
of electronic recoils using the difference in quenching effects between
the phonon/heat signal and either the ionization or the scintillation
signal. %
\begin{figure}
\includegraphics[%
  scale=0.45]{schembolo.epsi}\includegraphics[%
  scale=0.47]{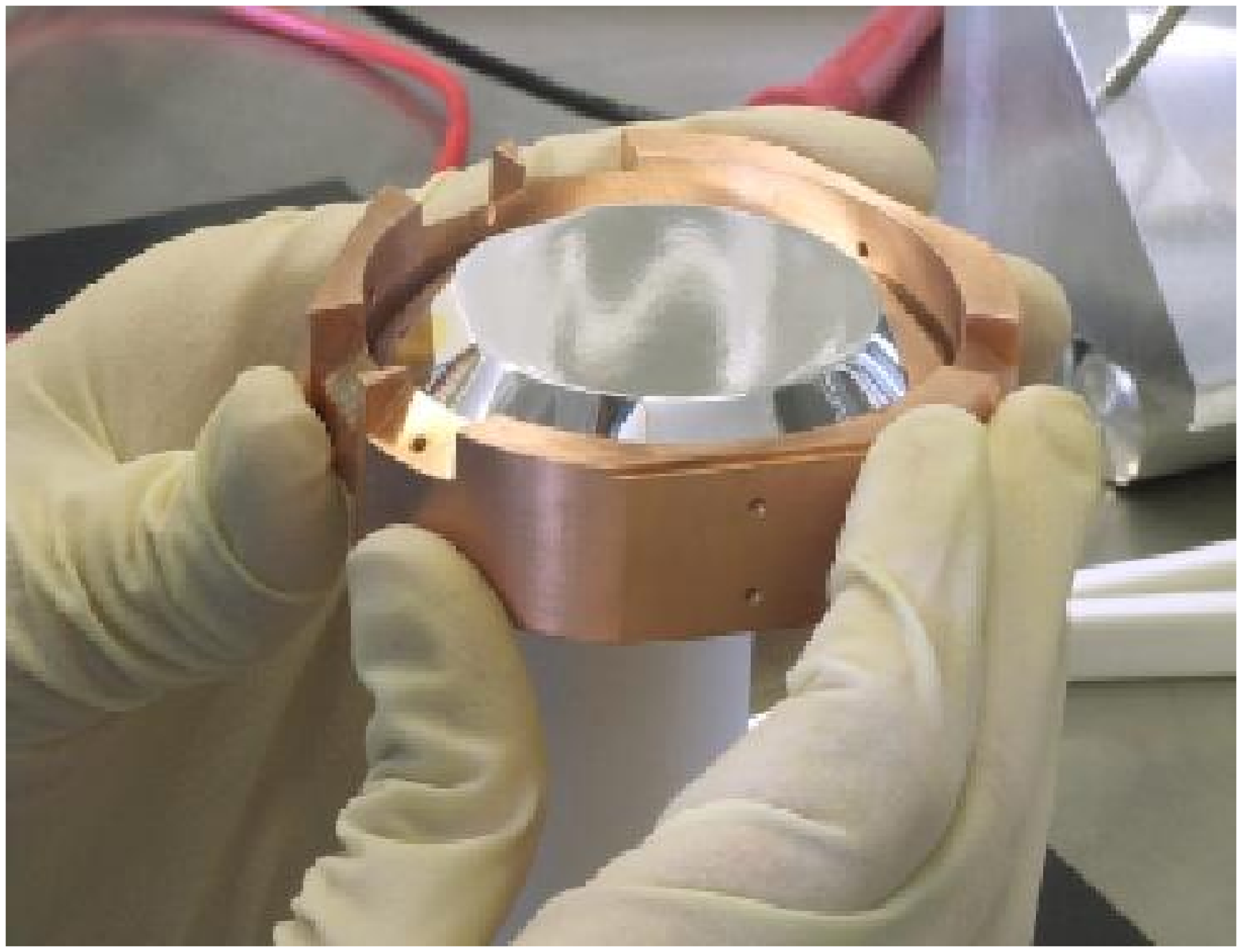}

\caption{\label{cap:Schembolo}Left: schematics of the heat-and-ionization
detector of the EDELWEISS collaboration (see text). Right: picture
of a detector in its copper support frame, before the wiring of the
electrode and the NTD heat sensor.}
\end{figure}
As an example, Fig.\ref{cap:Schembolo} shows the schematic view of
the heat-and-ionization of the EDELWEISS collaboration\cite{edw2002}.
An energy deposit in the detector will result in the creation of electron-holes
pairs in the semiconductor crystal, collected on Al-sputtered electrodes
polarized at a bias of a few Volts. In a few ms after the interaction,
the entire incident energy is thermalized in the detector, cooled
down to 17~mK in order to reduce the heat capacitance of the crystal
an produce a temperature increase of a few $\mu$K, measured with
a Neutron Transmutation Doped Ge thermistance glued to the side of
the detector. With the simultaneous measurement of the heat and ionization
signals for each event, one can deduce the true recoil energy and
the ionization quenching associated to it. %
\begin{figure}
\begin{center}\includegraphics[%
  scale=0.5]{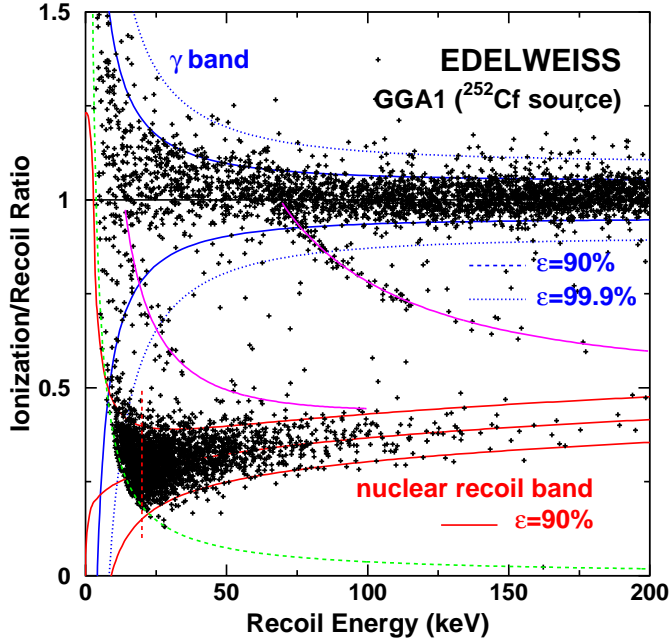}\end{center}

\caption{\label{cap:discring}Distribution of the the quenching factor (ratio
of the ionization signal to recoil energy) as a function of the recoil
energy recorded with a 320~g EDELWEISS detector exposed to a $^{252}$Cf
source emitting neutrons and photons. The full lines correspond to
the $\pm$1.645$\sigma$ bands (90\% efficiency) for electronic and
nuclear recoils and the dotted lines to the $\pm$3.29$\sigma$ band
(99.9\% efficiency) for electron recoils.}
\end{figure}
Fig.\ref{cap:discring} shows that this technique can discriminate
efficiently nuclear and electron recoils, down to low recoil energy.

The ZIP detectors developed by the CDMS collaboration are based on
the same principle, except that the heat sensor is replaced with a
thin film sensor able to detect phonons before their complete thermalization.
As discussed later on, the detection of this fast component leads
to the possibility to identify energy deposit close to the surface
by using the time evolution of the rise of the phonon signal.

The heat-and-ionization technique have come to a mature stage and
are now providing the best published sensitivities of $\sim$0.2 nuclear
recoil per kg$\cdot$d, achieved by the EDELWEISS\cite{edw2002} and
CDMS\cite{cdms1} cryogenic Ge detectors (3$\times$320~g and 4$\times$160~g,
respectively). More than 99.9\% of the electronic recoils are rejected
down to 15 and 10 keV, respectively. Since then the two experiments
have presented updated preliminary results.

EDELWEISS has increased its total exposure from 11.7 kg$\cdot$d to
62 kg$\cdot$d, and has reduced its threshold to 11 keV recoils. In
total, 3 events are observed in the critical 30 to 100 keV range.
More events (34) are observed at lower energy. The 90\% C.L. limits
on $\sigma_{n}$ deduced from the observed events is essentially the
same as those from the smaller published data set, as can be seen
on %
\begin{figure}
\begin{center}\includegraphics[%
  scale=0.5]{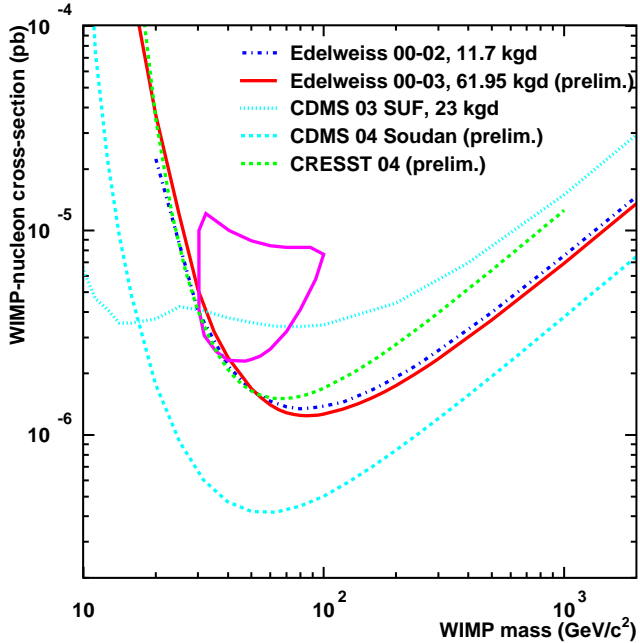}\end{center}

\caption{\label{cap:edwlim}90\% C.L. confidence limits on the spin-independent
WIMP-nucleon cross-section as a function of the WIMP mass for the
different cryogenics experiments described in the text.}
\end{figure}
Fig.\ref{cap:edwlim}. A coincidence between two detectors, with both
hits being identified as nuclear recoils, indicates that a residual
neutron background is likely present, and the WIMP limits, derived
with no background subtraction, are conservative. The installation
of the next phase of the experiment is now under way. It will involve
28 detectors, with 7 being equipped with NbSi thin film sensors, the
latter being able to use athermal phonon detection for the rejection
of the near-surface events that may yield deficient ionization signals.

Recently, CDMS\cite{cdms2} has presented preliminary results obtained
with four 150~g Ge detectors equipped with athermal phonon sensors,
operated at the SOUDAN underground site. After cuts on the timing
information to remove surface events, an exposure of 19.4 kg$\cdot$d
is obtained with at most one event observed in the 10-100 keV range.
The resulting 90\% C.L. limits, shown in Fig.\ref{cap:edwlim}, are
the first to go below the $10^{-6}$pb sensitivity.

Another exciting development is the preliminary results\cite{cresst2}
obtained by the CRESST collaboration with heat-and-scintillation detectors\cite{cresst}.
In their 300~g CaWO$_{4}$ detector, the ratio of the scintillation
signal to the heat signal provides a 99.9\% rejection of electron
recoils. In addition, their preliminary measurements of the relative
light yields of Ca, W and O recoils indicate that the light yield
for W recoils is significantly less than for Ca or O recoils. In this
detector one expects not two but three distinct populations: electron
recoils, O and Ca recoils (primarily due to neutron scattering) and
W recoils. Because of the $\mu^{2}A^{2}$ dependence, WIMPs are expected
to interact primarily with W nuclei, while neutrons will interact
relatively more often with O and Ca nuclei. The observation of no
events in the W recoils band in the interval from 12 to 40 keV in
a 10.5 kg$\cdot$d exposure yields the preliminary 90\% C.L. limit
shown on Fig.\ref{cap:edwlim}.

CDMS-II and CRESST-II are both pursuing their data taking with 10$^{-8}$pb
sensitivity goals, as for EDELWEISS-II, which will resume its operations
in 2005. For the 10$^{-10}$pb horizon, one-ton size cryogenics arrays
of detectors are being studied by both CDMS (CRYOARRAY\cite{cryoarray})
and CRESST-EDELWEISS (EURECA).

\subsection{Other techniques}

This review of the experimental techniques was focused on those that
have lead, in the past, or are currently leading the field of direct
dark matter searches in the context of the minimal supersymmetric
models. It is in no way exhaustive. As current experiments are more
than four order of magnitude away from a full coverage of the bulk
of supersymmetric predictions, the coming years may reveal that the
ultimate sensitivity can only be reached by detector techniques that
are now in a very early development stage. For example, detectors
sensitive to the recoil direction such as low-pressure Time Projection
Chamber (see e.g. Ref.\cite{drift}) may be essential for the exploration
of the kinematics of the WIMP flux on earth. Or it could be that the
WIMPs do not follow strictly the behavior suggested by the most common
forms of the supersymmetric models. For example, if the WIMP would
interact almost solely via spin-dependent interactions, detectors
made of nucleus with a large intrinsic spin, like those developed
by the PICASSO and SIMPLE collaborations\cite{picasso}, could play
a more important role.

\section{Conclusions}

The field of direct WIMP search is stimulating an intense detector
development effort aimed at achieving the sensitivity required for
the extremely low rates and low energy involved. Cryogenic detectors
with event-by-event identification of nuclear recoils have for now
taken the lead in this domain, but there is still a lot of development
in progress on the road to the 10$^{-8}$pb sensitivity of current
projects to the ultimate 10$^{-10}$pb sensitivity necessary to cover
most of the MSSM domain. In fine, only a combination of experimental
signatures, and thus likely of different detector techniques, will
produce a satisfying positive identification of the true nature of
the WIMP.

\section*{Acknowledgments}

Je remercie les organisateurs de l'école pour leur invitation et les
félicite pour leur succès.

\end{document}